\documentstyle[aps,pre,manuscript,epsf]{revtex}

\begin{document}

\title{``Single-cycle'' ionization effects in laser-matter interaction}
\author{Enrique Conejero Jarque\footnote[1]{On leave from Departamento de
F\'{\i}sica Aplicada, Universidad de Salamanca, Spain. E-mail
address enrikecj@gugu.usal.es} and Fulvio
Cornolti\footnote[2]{Also INFM, sezione A, Universit\'a di Pisa,
Italy}}

\address{Dipartimento di Fisica, Universit\'a di Pisa,
Piazza Torricelli 2, Pisa, Italy}

\author{Andrea Macchi\footnotemark[2] and Hartmut Ruhl\footnote[3]{Present address
Max-Born Institut, Max-Born str. 2a, 12489 Berlin, Germany}}

\address{Theoretical Quantum Electronics, Darmstadt University of
Technology, Darmstadt, Germany}

\maketitle

\begin{abstract}
We investigate numerically effects related to ``single-cycle''
ionization of dense matter by an ultra-short laser pulse. The
strongly non-adiabatic response of electrons leads to generation
of a MG steady magnetic field in laser-solid interaction. By using
two-beam interference, it is possible to create periodic density
structures able to trap light and to generate relativistic
ionization fronts.

\end{abstract}


\section{Introduction}
In the adiabatic field ionization regime, the ionization rate
grows sharply when the electric field approaches the barrier
suppression (BS) limit, i.e. when the laser intensity is high
enough that the electron in the ground state is able to
``classically'' escape the atomic potential barrier. The
ionization rate for such field strength may become higher than the
laser frequency and a regime in which most of the ionization is
produced within a single laser half-cycle is achievable.

Here we present a numerical study of some effects of ultrafast
ionization in the interaction of a short laser pulse with an
initially transparent dense medium. First, we will discuss the
generation of megagauss steady magnetic fields in the surface
``skin'' layer of ``solid'' targets, i.e. slabs of hydrogen atoms
with a number density close to that of a solid medium (Macchi et
al. 1999).

Second, we will describe effects related to the combination of
two-beam interference with ultrafast ionization. We will show how
it is possible to take advantage of this feature to create a
layered dielectric-conductor structure able to trap the electric
field, as well as a relativistic ionization front (Conejero Jarque
et al. 1999).

\section{Generation of steady magnetic fields}

The generation of steady currents and magnetic fields by ultrafast
ionization is due to the non-adiabatic nature of the response of
initially bound electrons to a strongly ramping laser field. Using
a following ``simple-man's'' model (SMM), very similar to the SMM
used in studies of above-threshold ionization and harmonic
generation in atoms, 
it can be shown that a single electron subject to an external
sinusoidal intense field can acquire a steady velocity (Macchi et
al. 1999)

\begin{equation} \label{eq:vsteady}
v_{st}=v_I-v_{qo}\sqrt{1-(E_T/E_{yo})^2},
\end{equation}

where $v_I$ is the ejection velocity of the electron,
$v_{qo}=eE_{y0}/m\omega$, $E_{yo}$ is the maximum field amplitude
and  $E_T$ is the field amplitude at the instant of ionization,
which will be close to the threshold field for barrier suppression
(for hydrogen, $E_T \approx 0.146 E_{au}= 7.45 \times 10^8 \mbox{
V cm}^{-1}$, being $E_{au}=5.1 \times 10^9 \mbox{ V cm}^{-1}$ the
atomic field unit).

If most of the electrons in the medium are ionized at the same
instant, as may happen with a pulse which sharply rises above
$E_T$, one gets a net steady current which in turn generates a
magnetic field. To obtain a larger current one may think to
``tune'' appropriately $E_T$ and $E_{y0}$. This is possible if the
ionization is no longer correlated with the oscillating field,
i.e., it is produced {\em independently} of the field itself, like
in the case studied by Wilks et al. (1988), in which a steady
magnetic field $B_{st} \approx E_{y0}$ can be obtained in a very
dense medium. For intense lasers ($I \geq 10^{18} \mbox{ W
cm}^{-2}$), such a magnetic field would get values exceeding $100
\mbox{ MG}$ and could explain (Teychenn\'{e} et al. 1998) the
experimental observation of high transparency of thin foil solid
targets to $30 \mbox{ fs}$, $3 \times 10^{18} \mbox{ W cm}^{-2}$
pulses (Giulietti et al., 1997). However, it is questionable
whether this high magnetic field may be obtained with superintense
laser pulses. In this case, in fact, the ``source pulse'' itself
ionizes the medium and thus this will impose a constraint on the
phase mismatch between the field and the velocity of the
electrons. We will show by numerical simulations that the steady
magnetic field exists but has values around $1 \mbox{ MG}$, being
therefore too weak to allow enhanced laser propagation.

\subsection{PIC simulations}
First we review the results of 1D3V PIC simulations with field
ionization included. We choose pulses with a ``$\sin^2$'' envelope
and with a ``square'' envelope. For all the PIC runs, the laser
frequency was $\omega_L=2 \times 10^{15} \mbox{ s}^{-1}$, close to
that of Nd and Ti:Sapphire lasers. The thickness of the target was
$0.09 \,\mu\mbox{m}$ and the density was $n_o=6.7 \times 10^{22}
\mbox{ cm}^{-3}$ ($\omega_{po}\simeq 7\omega_L$). For the
ionization rate we used a semi-empirical formula obtained from
atomic physics calculations
(Bauer and Mulser, 1999). 
The laser energy loss due to ionization is included introducing a
phenomenological ``polarization'' current (Rae and Burnett 1992,
Cornolti et al. 1998, Mulser et al. 1998).

\begin{figure}
\epsfxsize=12cm \epsfysize=5cm \centerline{\epsfbox{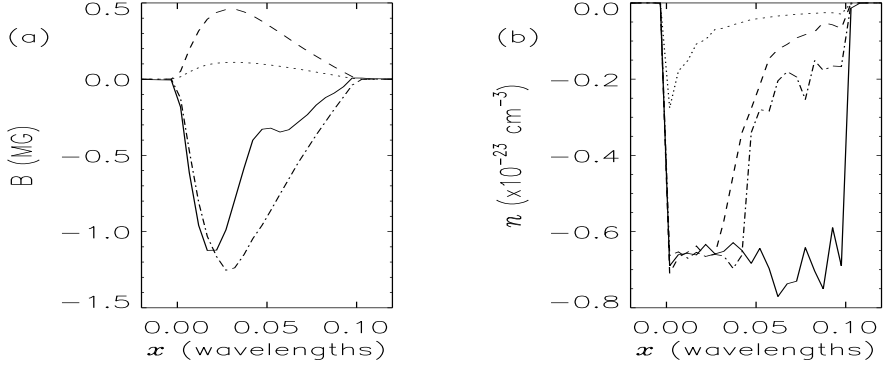}}
\caption[]{Spatial distribution of magnetic field (left) and
electron charge density (right) five cycles after the end of the
pulse, for ``$\sin^2$'' pulses of 0.1 a.u. (dotted line), 1 a.u.
(dashed line), 10 a.u. (solid line) maximum amplitude and a
``square'' pulse of 1 a.u. amplitude (dashed-dotted line). All the
pulses are 5 cycles long. The electric field atomic unit is
$E_{au}=5.1 \times 10^9 \mbox{ V cm}^{-1}$ (corresponding to
$I=3.5\times 10^{16}\mbox{ W cm}^{-2}$).} \label{Fig2}
\end{figure}

Fig.\ref{Fig2} shows the spatial profiles of the magnetic field
and the free electron density five cycles after the end of a five
cycles long ($\Delta t_L=15 \mbox{ fs}$) pulse, for three
different field intensities in the ``$\sin^2$'' shape case, and
for the square profile case at the intermediate intensity value.
The steady field is generated at the beginning of the interaction
and is always much weaker than the laser field, even for the most
intense case (corresponding to an intensity of $3.5\times
10^{18}\mbox{W cm}^{-2})$; its sign varies according to the phase
of the laser half cycle where most of the ionization occurs. The
ionization at the left boundary is nearly instantaneous; however,
even if the target is only $0.1 \lambda$ thick, it is not ionized
over its whole thickness due to instantaneous screening, except
for the maximum intensity case.

The fact that the produced magnetic field is much less than
expected may be attributed to the instantaneous screening of the
EM wave due to the ultrafast ionization. In fact, it is too weak
to affect self-consistently the refractive index and as a
consequence it cannot lead to magnetically induced transparency as
hypothesized by Teychenn\'{e} et al. (1998).

\subsection{Boltzmann simulations}

To yield a further insight into the magnetic field generated by
ultrafast ionization we look at the results of 1D and 2D Boltzmann
simulations. This corresponds to the ``direct'' numerical solution
of the Boltzmann equation for the electron
distribution function $f_e=f_e ({\bf x},{\bf v},t)$, over a phase space
grid:
\begin{equation}
\partial_t f_e+{\bf v}\cdot\nabla f_e
-\frac{e{\bf E}}{m}\cdot  \partial_{\mbox{{\bf v}}} f_e= \nu_I(E)
n_a({\bf x},t)g({\bf v};{\bf E}({\bf x},t)).
\label{eq:ioniz_kinet}
\end{equation}
Here $n_a$ is the density of neutral atoms (supposed at rest for
simplicity) and $\nu_I$ is the ionization rate. The term $g({\bf
v};{\bf E})$ gives the ``instantaneous'' distribution of the just
ionized electrons, which is supposed to be known from atomic
physics. A semiclassical picture which allows to define and
evaluate $g({\bf v};{\bf E})$ was given by Cornolti et al.
(1998). 
With respect to PIC simulations, the Boltzmann approach has the
disadvantage of larger memory requirements, but the advantages of
reduced numerical noise and the possibility to take into account
the full kinetic distribution of the ionized electrons.

\begin{figure}
\unitlength1cm
\begin{center}
\begin{picture}(14,6)
\epsfxsize=6cm \epsfysize=6cm \put(0,0){\epsfbox{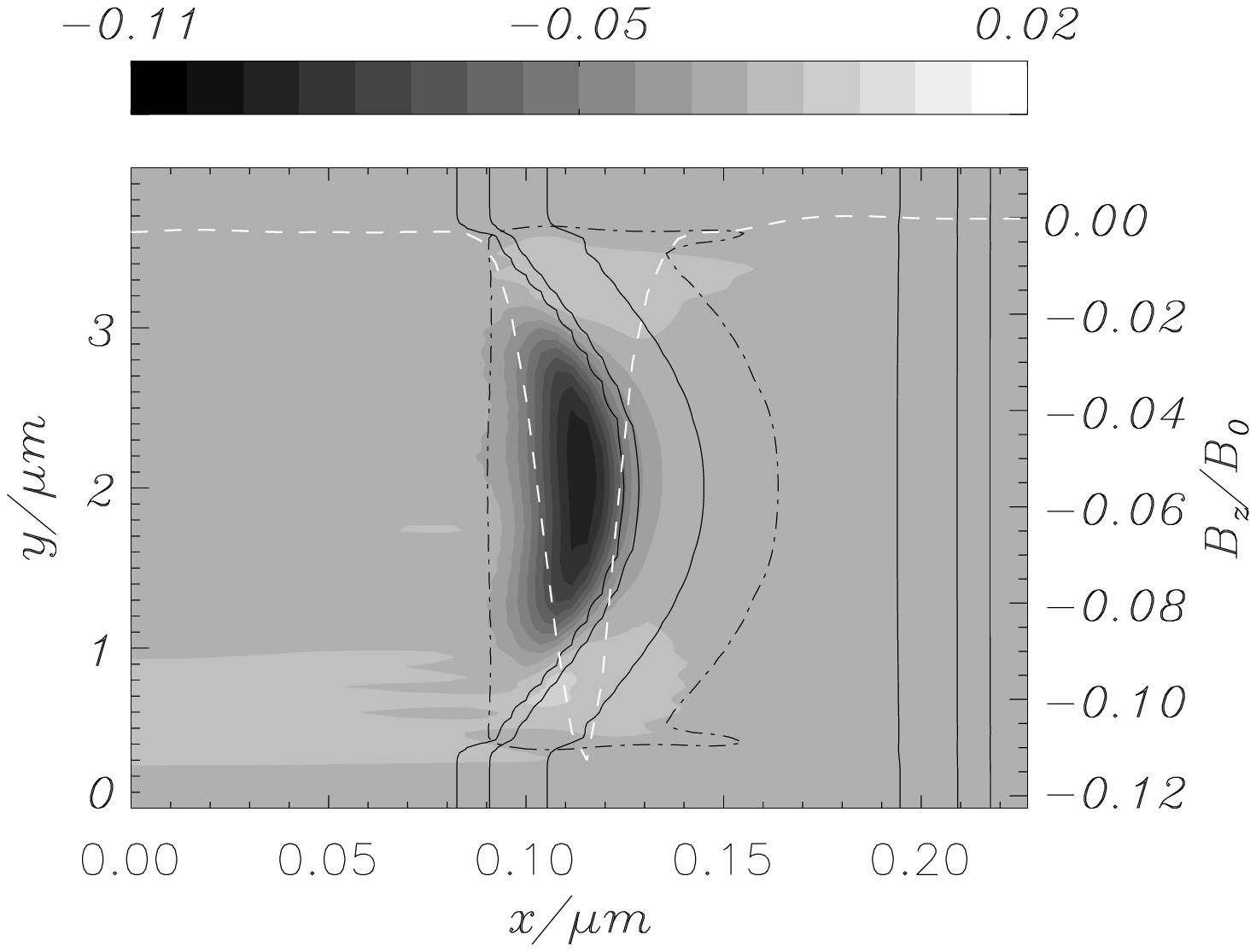}}
\put(-0.5,5){\large (a)} \epsfxsize=6cm \epsfysize=6cm
\put(7,0){\epsfbox{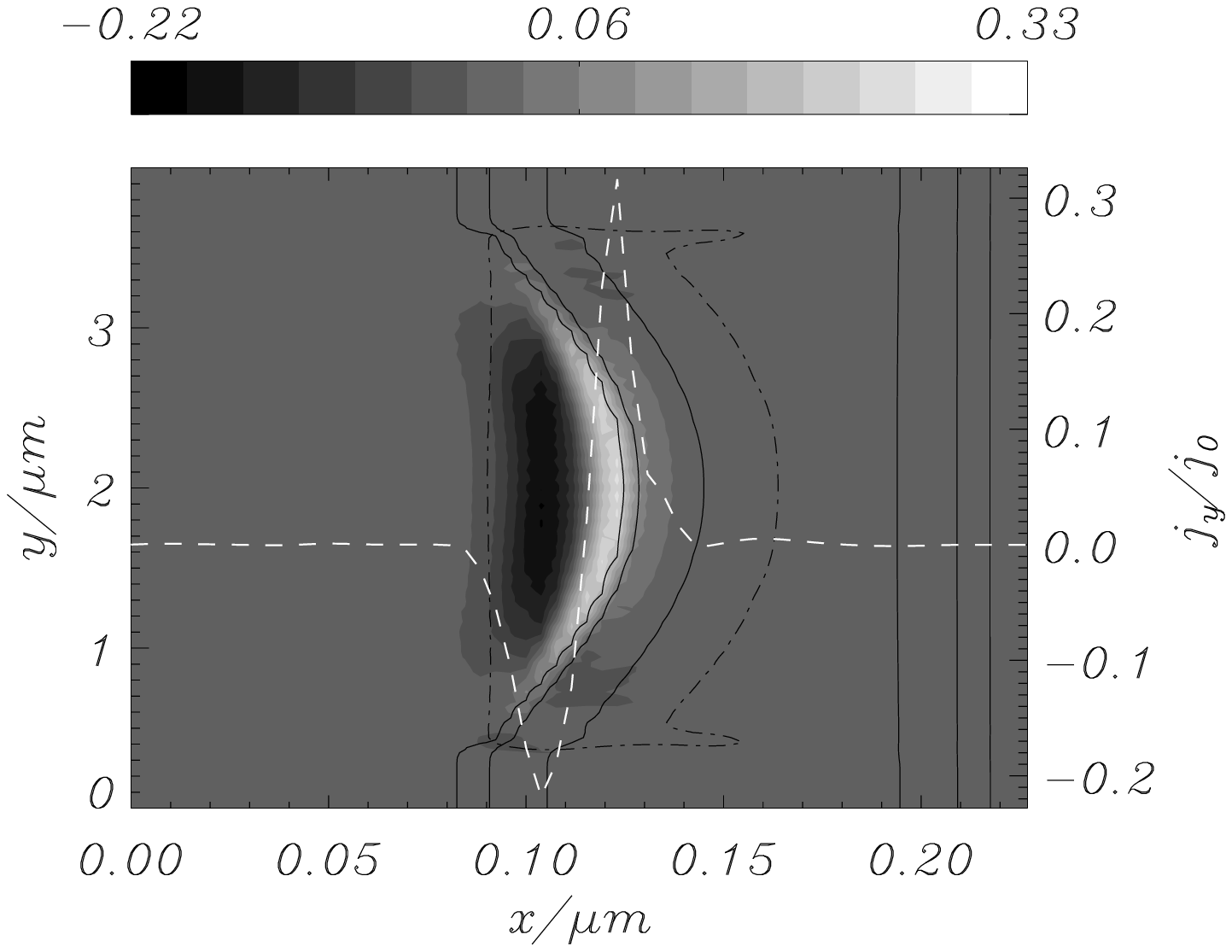}} \put(6.5,5){\large (b)}
\end{picture}
\end{center}
\caption[]{Grayscale contours of the magnetic field $B_z$ (a) and
the current density $j_y$ (b) five laser cycles after the laser
pulse end, for a 2D2V Boltzmann simulation. The dashed line in (a)
and (b) give $B_z/B_o$ and $j_y/j_o$, respectively, along $x=2
\mu{m}$. The parameters $B_o=27.7 \mbox{ MG}$, $j_o=2.2 \times
10^{22}$ c.g.s. units. The solid lines give neutral density
contours. The dashed-dotted lines mark the critical density
surface. Simulation parameters $I=10^{16} \mbox{W cm}^{-2}$,
$\lambda=0.25 \mu\mbox{m}$, $n_e/n_c=12.5$.} \label{Fig3}
\end{figure}

We first look at 2D2V Boltzmann simulations. We take a $0.25
\,\mu\mbox{m}$, $10^{16} \mbox{ W cm}^{-2}$ laser pulse impinging
on a solid hydrogen target with number density $2 \times 10^{23}
\mbox{ cm}^{-3}=12.5 n_{c}$, and thickness $0.1 \,\mu\mbox{m}$.
The time envelope of the laser pulse is Gaussian with a FWHM
duration of 2 cycles. The laser spot is also Gaussian with a FWHM
of $2 \,\mu\mbox{m}$. Fig.\ref{Fig3} (a) shows the magnetic field
and the density contours after the end of the laser pulse. The
steady magnetic field has constant (negative) sign over its
extension. Its maximum intensity is about $3 \mbox{ MG}$.
Fig.\ref{Fig3} (b) shows the electron current density $j_y$ at the
same time of the right plot of fig.\ref{Fig3} (a).

\begin{figure}
\unitlength1.0cm
\begin{center}
\begin{picture}(12,5)
\epsfxsize=6cm \epsfysize=5cm \put(0,0){\epsfbox{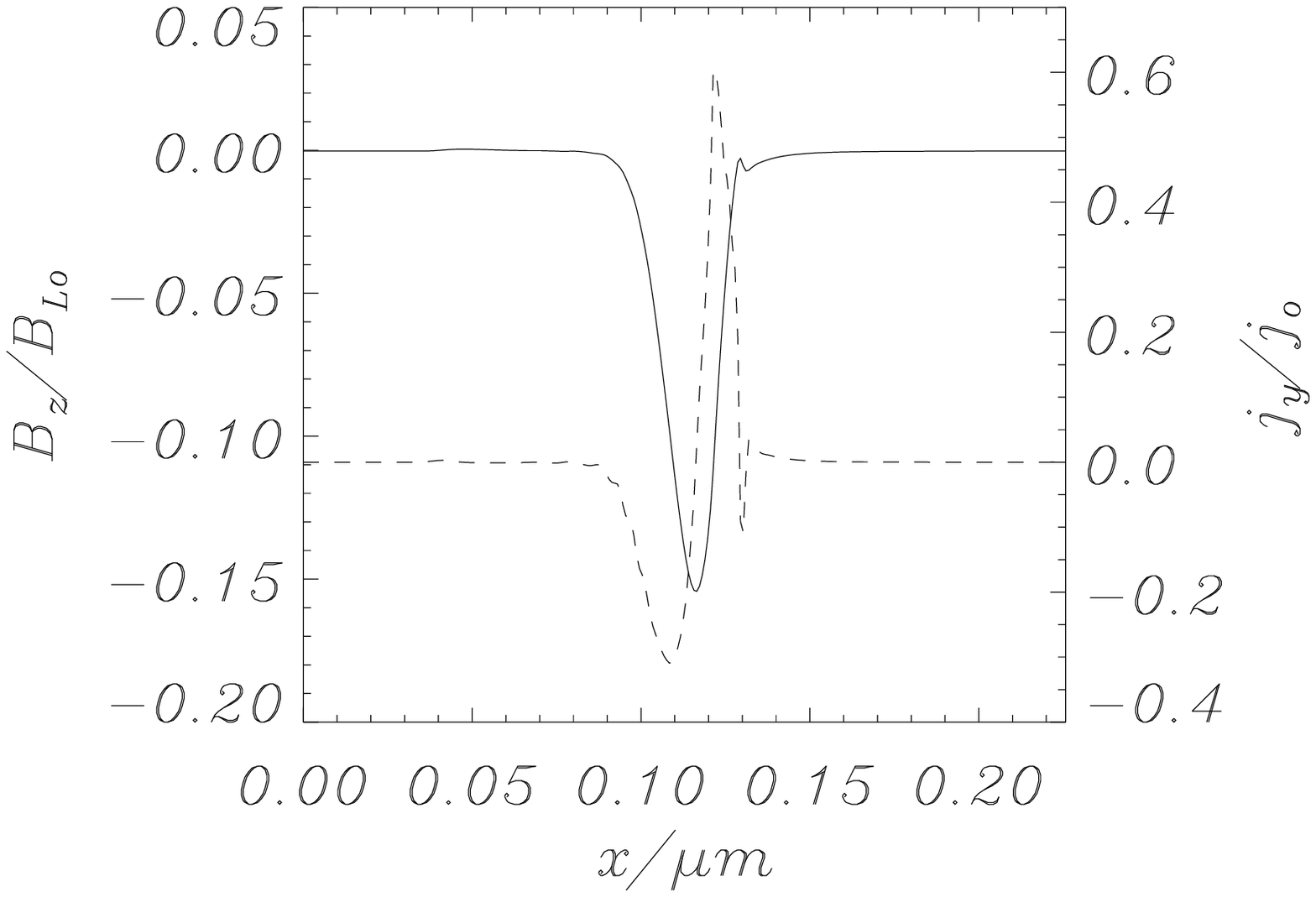}}
\epsfxsize=6cm \epsfysize=5cm \put(6,0){\epsfbox{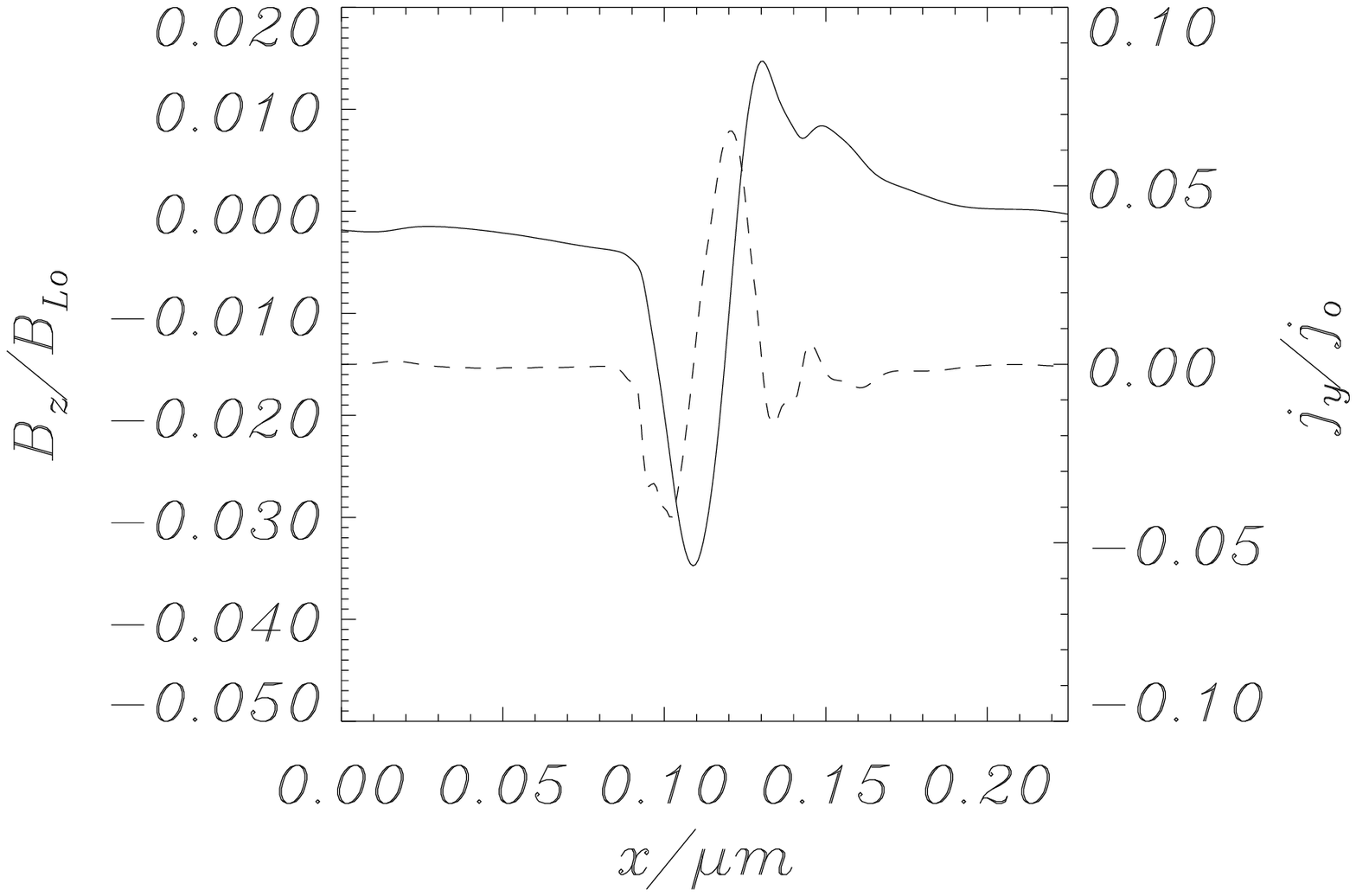}}
\put(-0.5,4.5){\large (a)} \put(5.5,4.5){\large (b)}
\end{picture}
\end{center}
\caption{Profiles of the steady magnetic field $B_z$ (solid) and
the current $j_y$ (dashed) in 1D Boltzmann simulations. The
parameters common to (a) and (b) are $\lambda=0.25 \,\mu\mbox{m}$
and $n_e/n_c=12.5$. In the case (a) the atomic parameters are
those of an hydrogenlike atom with $Z=1$, and $I=10^{16} \mbox{W
cm}^{-2}$, $B_{o}=27.7 \mbox{ MG}$, $j_o=2.2 \times 10^{22}$
c.g.s. units. In the case (b) $Z=2$ and laser parameters are
scaled accordingly to  ${\bf x}\rightarrow Z{\bf x}$,
$t\rightarrow Z^2 t$, $\omega \rightarrow Z^{-2}\omega$, ${\bf E}
\rightarrow Z^{-3}{\bf E}$; $I=6.4 \times 10^{17} \mbox{W
cm}^{-2}$, $B_{o}=50.6 \mbox{ MG}$, $j_o=7 \times 10^{22}$ c.g.s.
units.}\label{Fig4}
\end{figure}

Among the parameters of our simulations, the magnetic field
appears to be most sensitive to the temporal profile of the laser
pulse, achieving its maximum value for a square pulse with zero
risetime. In Fig.\ref{Fig4} (a) we show the results of a 1D
Boltzmann simulations for a square pulse with $I=10^{16} \mbox{W
cm}^{-2}$, $\lambda=0.25 \,\mu\mbox{m}$, and a target with
$n_e/n_c=12.5$. The current density is $j_y \sim 10^{22}$ c.g.s.
units and extends over a distance comparable to $d_p \simeq 1.2
\times 10^{-2}\,\mu\mbox{m}$. The maximum magnetic field is
consistent with Ampere's law, which gives $B_{st} \sim 4 \pi j_y
d_p/c \simeq 5 \mbox{ MG}$. Assuming a density $n_e \simeq n_o =
2.2 \times 10^{23} \mbox{ cm}^{-3}$ for the electrons which are
instantaneously ionized, one gets a steady velocity $v_{st} \simeq
j_y/en_e \simeq 10^8 \mbox{ cm s}^{-1}$. This value is lower than
the ejection velocity for hydrogen $v_I \simeq 2 \times 10^8
\mbox{ cm s}^{-1}$. This suggests that effects such as screening,
nonzero ionization time, and velocity statistics act to keep the
steady current well below the values that one may estimate
according to the SMM, eq.(\ref{eq:vsteady}).

Both laser and target parameters where varied in simulations in order to
investigate the scaling of the magnetic field with them.
As an example, Fig.\ref{Fig4} (b)
shows the results of a simulation for a target of hydrogenic ions
with density and thickness identical
to Fig.\ref{Fig4} (a), but where we assumed a nuclear charge $Z=2$
and scaled the atomic parameters accordingly to ${\bf x}\rightarrow
Z{\bf x}$, $t\rightarrow Z^2 t$, $\omega \rightarrow
Z^{-2}\omega$, ${\bf E} \rightarrow Z^{-3}{\bf E}$. In order to
have the ionization threshold to be exceeded at the same instant,
the laser pulse had the same envelope and frequency but the
intensity was scaled by $Z^6$. With respect to the $Z=1$ case, we obtain
a steady field with {\em lower} peak amplitude which assumes both
positive and negative values.

We also performed 2D Boltzmann simulations for a pulse obliquely
incident at $15^o$ on the target. The preliminary results show
that the magnetic field is much lower in this case. Therefore it
appears that the steady magnetic field is sensitive to the
interaction geometry. In any case, the oblique incidence results
further confirm the conclusion that no magnetic field capable to
affect the transmission through the target is generated.

\section{Optical microcavities and
ionization fronts}

\subsection{The model}

In this section, we study effects related to two beam-interference
in one spatial dimension and for wavelengths in the infrared and
optical range. In our numerical experiment, a one-dimensional
interference pattern is generated via  an appropriate ``target
manufacturing'': the idea is to place a reflecting mirror on the
rear side of the target, the one opposite to the laser. Such a
mirror might be easily produced by a metallic coating on a glass
or plastic target. Taking a laser pulse with peak intensity
between $I_T/2$ and $I_T$, being $I_T$ the ``threshold'' value for
ionization, a plasma is produced in the target bulk  around the
maxima of interference pattern produced by the incident wave and
the wave reflected at the rear mirror.

Since in this regime we deal only with moderate laser intensities,
we may use a simple one-dimensional fluid model based on
continuity, current and wave equations for an ionizing medium,
originally proposed by Brunel (1990), modified by the inclusion of
the polarization current. More details about the model and its
validity can be found in Cornolti et al. (1998) and Conejero
Jarque et al. (1999).

\subsection{Generation of layered plasmas}

We first consider a target with thickness $L=2 \pi \lambda$, being
$\lambda = 0.8 \,\mu\mbox{m}$, and density $n_o=10n_c$. The laser
pulse has a $\sin^2$-shaped envelope with a duration of $80 \mbox{
fs}$ ($30$ cycles) and a peak intensity $I=1.8 \times 10^{14}
\mbox{ W cm}^{-2}$. The target parameters are chosen to simulate a
thin foil solid slab and it is enough to take the density as low
as $10n_c$ since the maximum electron density always remains much
lower than this value.

\begin{figure}
\epsfxsize=8cm \epsfysize=7cm\centerline{\epsfbox{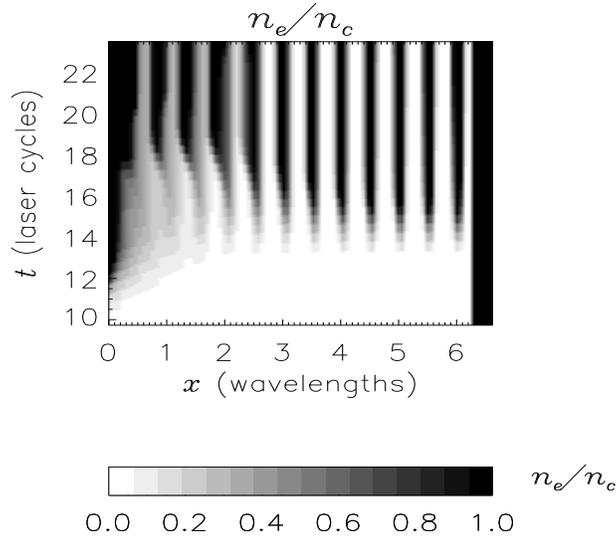}}
\vskip1cm \caption{Grayscale contourplot of free electron density
$n_e(x,t)$ for a ``solid'' hydrogen target with a reflecting
``metal'' layer on the rear face (thick solid line). The pulse
parameters are $I=1.8 \times 10^{14} \mbox{ W cm}^{-2}$, $\lambda=
0.8 \,\mu\mbox{m}$, $\Delta t_L=30(2\pi/\omega_L) \simeq 80 \mbox{
fs}$ (``$\sin^2$'' envelope). The target parameters are $L=2 \pi
\lambda$, $n_o=10 n_c =1.1 \times 10^{22} \mbox{ cm}^{-3}$.}
\label{fig1_n}
\end{figure}

The electron density vs. space and time is shown in
Fig.\ref{fig1_n}. A clear layered density pattern with a spatial
periodicity close to $\lambda/2$ is produced along nearly all the
slab. The layers of overdense plasma are produced near the maxima
of the interference pattern. These maxima appear at close times
because of the effect of the smooth envelope of the laser pulse.
The resulting quasi-periodic structure of the refractive index has
in principle some similarities with the widely studied
semiconductor microcavities and photonic band-gap materials (see
reviews by Burstein and Weisbuch (1993) and by
Skolnick et al. (1998)).

\subsection{Optical microcavities}
\begin{figure}
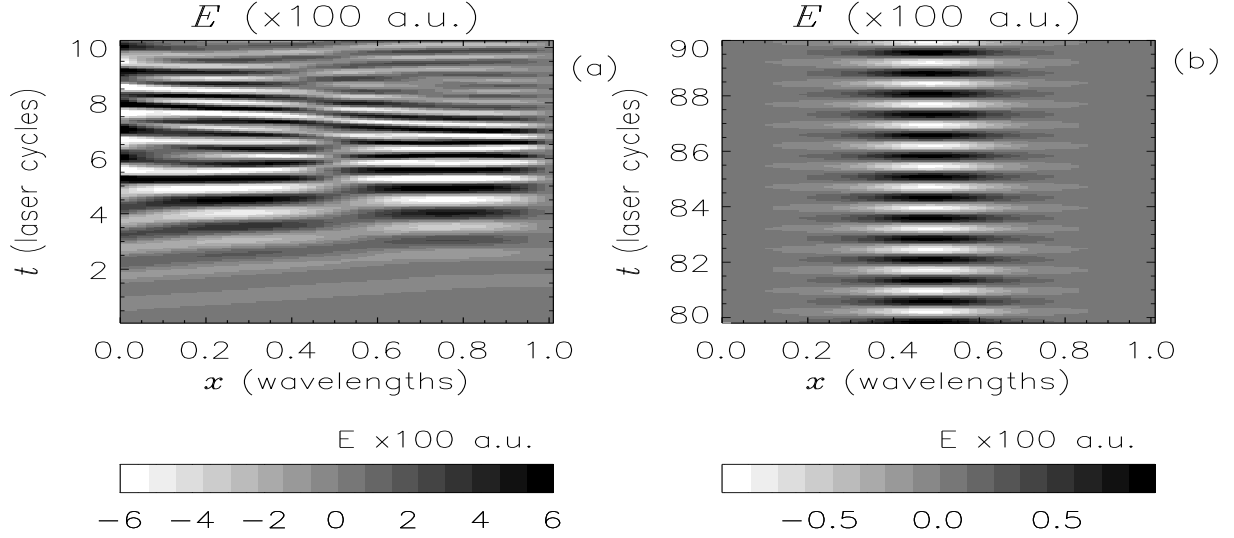

\centerline{\epsfxsize=8cm\epsfysize=7cm\epsfbox{fig3a.epsi}
\epsfxsize=8cm\epsfysize=7cm\epsfbox{fig3b.epsi}} \vskip1cm
\caption{Evolution of the electric field inside the plasma slab
during the interaction with the incident pulse (a) and 80 cycles
later (b). The pulse parameters are $I=1.8 \times 10^{14} \mbox{ W
cm}^{-2}$, $\lambda= 10.6 \,\mu{m}$, $\Delta t_L=15(2\pi/\omega_L)
\simeq 530 \mbox{ fs}$ (``$\sin^2$'' envelope). The target
parameters are $L=\lambda$, $n_o=5 n_c = 5\times 10^{19} \mbox{
cm}^{-3}$.} \label{fig3_n}
\end{figure}

Since the density in the plasma layers is overcritical, and the
layers are created in a time shorter than a laser halfcycle, the
portions of the standing wave between adjacent intensity maxima
may be ``trapped'' into the cavity formed by the two neighboring
layers. This trapping effect is best seen in the case of a CO$_2$
pulse impinging over a gas target with $L= \lambda=10.6
\,\mu\mbox{m}$ and $n_o=5n_c \simeq 5\times 10^{19}
\mbox{cm}^{-3}$. For this target, two plasma layers are produced
around the positions $x=0.25 \lambda, x=0.75\lambda$. Fig.
\ref{fig3_n} shows the map of the electric field at early (a) and
later (b) times, showing the generation of the constructive
interference pattern which yields the layered ionization (a), and
the subsequent trapping of the field which remains in the cavity
at times longer than the incident pulse duration (b). The
non-ionized regions between density layers clearly act as optical
microcavities.

Since the microcavity length is $L_c \leq \lambda/2$, light must
have an upshifted wavevector $k' \geq k$ in order to persist
inside the cavity. This implies also upshift of the laser
frequency with $\omega'_L \geq \omega_L$ as seen in
Fig.\ref{fig3_n}(b). The upshift decreases the critical density
value for the trapped radiation and therefore wavelengths much
shorter than $\lambda$ escape from the cavity. Due to the small
fraction of light that tunnels out of the cavity one observes
radiation emission from the target for a time much longer than the
pulse duration. Both the frequency upshift and the pulse
lengthening may provide experimental diagnostics for microcavity
generation. The lifetime of the cavities is ultimately limited by
processes such as recombination, which however should appear on
times much longer than the pulse duration of a few tens of
femtoseconds that are considered here and are available in the
laboratory.

\subsection{Ionization fronts}

As already shown, in our model target ionization is produced
around the maxima of the ``standing'' wave which is generated due
to the reflection at the rear mirror. However, since ionization is
instantaneous on the laser period timescale, it is produced as
soon as the wave reflected at the rear mirror travels backwards
and builds up the standing wave by interference. Therefore, a
backward propagating ionization front is generated, as seen in
Fig.\ref{fig1_n}. The density at the front exceeds the critical
density. This feature is not obtained for a single pulse impinging
on a dense target, since it undergoes immediate self-reflection
and penetrates only in the ``skin''surface layer (Macchi et al.
1999).

\begin{figure}
\unitlength1.0cm \epsfxsize=8cm
\epsfysize=7cm\centerline{\epsfbox{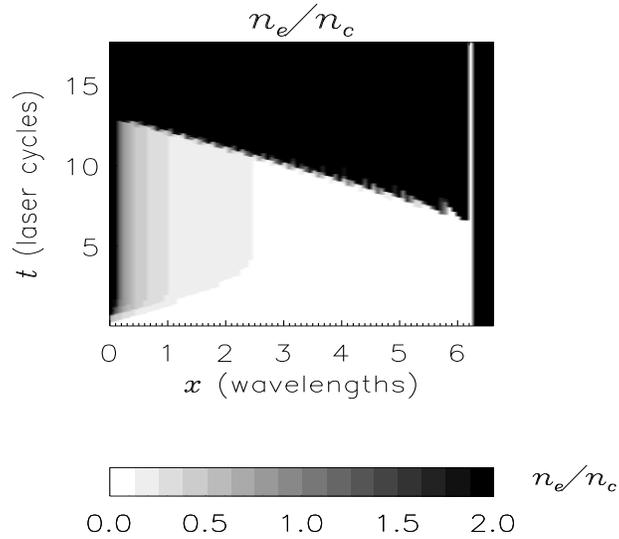}} \vskip1cm
\caption{Grayscale contourplot of $n_e(x,t)$ for a hydrogen
``gaseous'' target with a reflecting ``metal'' layer on the right
boundary. The pulse has square envelope and parameters $I=1.8
\times 10^{14} \mbox{ W cm}^{-2}$, $\lambda= 10.6 \,\mu{m}$,
$\Delta t_L=15(2\pi/\omega_L) \simeq 530 \mbox{ fs}$ Target
parameters are $L=2\pi\lambda$, $n_e=4 n_c =4 \times 10^{19}
\mbox{ cm}^{-3}$.} \label{fig8_n}
\end{figure}

An example of ``overdense'' ionization front is obtained in the
case of a CO$_2$ square pulse 15 cycles long impinging over a
target with $n_e=4n_c$. The $n_e(x,t)$ contour plot is shown in
Fig.\ref{fig8_n}. The ionized layers merge into a more homogeneous
distribution and a ``continuous'' ionization front appears. The
merging appears because the time- and space- modulated refractive
index perturbs the reflected wave substantially, leading to
broadening of interference maxima. The velocity of the front in
Fig.\ref{fig8_n} is near to, or even  exceeds at some times that
of light. This is clearly not a physical ``moving mirror'' with a
velocity greater than $c$, but a reflective surface which is
created apparently with such velocity due to a space-time phase
effect.

\section*{Acknowledgments}
We acknowledge the scientific contributions of D. Bauer and L.
Plaja as well as their suggestions.  Discussions with G. La Rocca,
R. Colombelli, L. Roso, and V. Malyshev are also greatly
acknowledged. This work has been supported by the European
Commission through the TMR networks SILASI, contract No.
ERBFMRX-CT96-0043, and GAUSEX, contract. No. ERBFMRX-CT96-0080.
E.C.J. also acknowledges support from the Junta de Castilla y
Le\'on (under grant SA56/99).

\section*{References}
\noindent BAUER, D. 1997 Phys. Rev. A {\bf 55}, 2180.

\noindent BAUER, D. \& MULSER, P. 1999 Phys. Rev. A {\bf 59}, 569.

\noindent BRUNEL, F. 1990 J. Opt. Soc. Am. B {\bf 7}, 521.

\noindent BURSTEIN, E. \& WEISBUCH, C., eds. 1993 {\em Confined
Electrons and Photons. New Physics and Applications} (NATO ASI
Series B: Physics, vol.340, Plenum Press, New York, 1993).

\noindent CONEJERO JARQUE, E., CORNOLTI, F. \& MACCHI, A. 2000 J.
Phys. B: At. Mol. and Opt. Phys. {\bf 33}, 1.

\noindent CORNOLTI, F., MACCHI, A. \& CONEJERO JARQUE, E. 1998 in
{\em Superstrong Fields in Plasmas}, Proceedings of the First
International Conference (Varenna, Italy, 1997), edited by M.
Lontano {\em et al.}, AIP Conf. Proc. No. {\bf 426} (AIP, New
York, 1998), p.55.

\noindent GIULIETTI, D., GIZZI, L.A., GIULIETTI, A., MACCHI, A.,
TEYCHENNE, D., CHESSA, P., ROUSSE, A., CHERIAUX, G., CHAMBARET,
J.P. \& DARPENTIGNY, G. 1997 Phys. Rev. Lett. {\bf 79}, 3194.

\noindent MACCHI, A., CONEJERO JARQUE, E., BAUER, D., CORNOLTI, F.
\& PLAJA, L. 1999 Phys. Rev. E {\bf 59}, R36.

\noindent MULSER, P., CORNOLTI, F. \& BAUER, D. 1998 Phys. of
Plasmas {\bf 5}, 4466.

\noindent RAE, S. C. \& BURNETT, K. 1992 Phys. Rev. A {\bf 46},
1084.

\noindent SKOLNICK, M. S., FISHER, T. A. \& WHITTAKER D. M. 1998
Semicond. Sci. Technol. {\bf 13}, 645.

\noindent TEYCHENN\'E, D., GIULIETTI, D., GIULIETTI, A. \& GIZZI,
L. A. 1998 Phys. Rev. {\bf E58}, R1245.

\noindent WILKS, S. C., DAWSON, J. M. \& MORI, W. B. 1988 Phys.
Rev. Lett. {\bf 61}, 337.

\newpage

\newpage

 \newpage

\end{document}